\documentclass[prd,superscriptaddress]{revtex4}
\usepackage[dvips]{graphicx}
\usepackage{color}
\usepackage{amsmath}
\usepackage[dvips]{graphicx}
\begin{document}

\title{Brown-York Energy in Spacetimes with Horizon}

\author{Jarmo M\"akel\"a} 

\email[Electronic address: ]{jarmo.makela@vamk.fi}  
\affiliation{Vaasa University of Applied Sciences, Wolffintie 30, 65200 Vaasa, Finland}

\begin{abstract}

We obtain a simple relationship between the change in the Brown-York energy inside of a closed two-surface just outside of a horizon of spacetime, and the change in the area of that two-surface.

\end{abstract}

\maketitle

Ever since the publication of Einstein's general relativity more than hundred years ago physicists have introduced several concepts of energy in curved spacetime. One of them is the {\it Brown-York energy}, which is defined in static, asymptotically flat spacetimes as: \cite{yy}
\begin{equation}
E_{BY} := -\frac{1}{8\pi}\oint_S(k - k_0)\,dS.
\end{equation}
In this definition $S$ is a closed, two-dimensional surface embedded into a spacelike hypersurface of spacetime, where the time coordinate $t=constant$. $dS$ denotes the area element on that two-surface, and we have integrated over the two-surface. $k$ is the trace of the exterior curvature tensor on the two-surface, and $k_0$ is the trace of the exterior curvature tensor in the limit, where the two-surface has been carried to the spatial infinity.

   The Brown-York energy may be understood as the total energy inside of the closed two-surface $S$. In this paper we shall show that when the two-surface $S$ is just outside of a closed, compact {\it horizon} of spacetime, the Brown-York energy has a very interesting property: If the proper acceleration vector field $a^\mu$ of an observer at rest on the two-surface $S$ has a constant norm $a$, is perpendicular to  $S$ at every point of $S$, and we change both the geometry of spacetime and the two-surface $S$ such that the proper acceleration $a$ is kept unchanged, then between the change $dE_{BY}$ in the Brown-York energy $E_{BY}$ and the change $dA$ in the area $A$ of the two-surface $S$ there is the relationship:
\begin{equation}
dE_{BY} = \frac{a}{8\pi}dA.
\end{equation}
In Ref. \cite{kaa} the validity of Eq. (2) was proved for the Schwarzschild spacetime, in Ref. \cite{koo} for the general, sphere symmetric spacetime with horizon, and in Ref. \cite{nee} for the Kerr-Newman spacetime. Since the proper acceleration is a constant, Eq. (2) suggests that we may consider the quantity
\begin{equation}
E = \frac{a}{8\pi}A
\end{equation}
as the total energy inside of the surface $S$ from the point of view of an observer with constant proper acceleration $a$ just outside of a horizon of spacetime.  Indeed, the energy of Eq. (3) was used as the energy of the gravitational field in Refs. \cite{kaa,koo,nee,vii,kuu,seite1}. A closely related energy concept has been introduced in Ref. \cite{seite}. 

As the starting point of the general derivation of Eq. (2), which is still lacking, we use the properties of the Rindler spacetime. As it is well known, Rindler spacetime is just two-dimensional, flat spacetime equipped with the so-called Rindler coordinates $t$ and $x$ such that the line element takes the form: \cite{kasi}
\begin{equation}
ds^2 = -x^2\,dt^2 + dx^2.
\end{equation}
In this equation the spatial coordinate $x$ is related to the proper acceleration $a$ of the observer such that
\begin{equation}
x = \frac{1}{a}.
\end{equation}
In other words, an observer with constant $x$ has constant proper acceleration $a$. The Rindler coordinates $t$ and $x$ are related to the flat Minkowski coordinates $T$ and $X$ such that:
\begin{subequations}
\begin{eqnarray}
t &=& \tanh^{-1}\left(\frac{T}{X}\right),\\
x &=& \sqrt{X^2 - T^2}.
\end{eqnarray}
\end{subequations}
The line element (4) is singular, when $x=0$, and Eq. (6b) implies that
\begin{equation}
T = \pm X,
\end{equation}
when $x=0$. The lines $T = \pm X$ constitute the so-called {\it Rindler horizon} from the point of view of the accelerating observer. According to Eq. (5) the proper acceleration $a$ of the observer tends to infinity at the Rindler horizon.

   We now proceed from the two-dimensional flat Rindler spacetime to a four-dimensional curved spacetime equipped with a horizon. As the first step we define a new spatial coordinate 
\begin{equation}
y := x^2,
\end{equation}
and Eq. (4) takes the form:
\begin{equation}
ds^2 = -y\,dt^2 + \frac{1}{4y}\,dy^2.
\end{equation}
We consider a spacetime metric
\begin{equation}
ds^2 = -y\,dt^2 + \frac{1}{4y}\,dy^2 + q_{jk}\,d\chi^j\,d\chi^k,
\end{equation}
where $j, k = 2,3$. In this equation $q_{jk}$ is the metric tensor induced on a spacelike, closed two-surface $S$ embedded into the spacelike hypersurface of spacetime, where the time coordinate $t=constant$. $\chi^2$ and $\chi^3$ are the coordinates on that two-surface. In what follows, we shall assume that $y$ is an injective, increasing function of coordinate $\lambda$, and an arbitrary function of $n$ parameters $\alpha^1,\alpha^2,\dots,\alpha^n$. In other words,
\begin{equation}
y = y(\lambda; \alpha^1,\alpha^2,\dots,\alpha^n).
\end{equation}
In contrast, the metric tensor $q_{jk}$ is assumed to be a function of $\lambda$ and the coordinates $\chi^2$ and $\chi^3$ {\it i. e.}
\begin{equation}
q_{jk} = q_{jk}(\lambda,\chi^2,\chi^3).
\end{equation}
Hence we find that if we denote:
\begin{equation}
y' := \frac{\partial y}{\partial\lambda},
\end{equation}
Eq. (10) takes the form:
\begin{equation}
ds^2 = -y\,dt^2 + \frac{1}{4y}(y')^2\,d\lambda^2 + q_{jk}\,d\chi^j\,d\chi^k.
\end{equation}
The coordinate $\lambda$ determines the location of the two-surface $S$ in spacetime, whereas the parameters $\alpha^s$ $(s=1,2,\dots,n)$ determine the overall geometrical properties of spacetime. For instance, if spacetime contains a black hole, the parameters may be, for example, the mass and the electric charge of the hole. Since the metric tensor $q_{jk}$ is a function of $\lambda$, the proper acceleration $a$ is everywhere a constant on the two-surface $S$, and it is perpendicular to $S$.  Spacetime equipped with the metric (14) has the horizon, when $y=0$. We choose the coordinate $\lambda$ such that
\begin{equation}
y(\lambda_0) = 0,
\end{equation}
and hence $\lambda=\lambda_0$ at the horizon. Since $y$ is assumed to be an increasing, injective function of $\lambda$, we must have
\begin{equation}
y'(\lambda_0) >0.
\end{equation}

In general, the proper acceleration vector field of an observer is
\begin{equation}
a^\mu = u^\sigma u^\mu_{;\sigma},
\end{equation}
where $u^\mu$ is the future pointing unit tangent vector field of the world line of the observer. For an observer at rest with respect to the spatial coordinates $\lambda$, $\chi^2$ and $\chi^3$ the only non-zero component of $u^\mu$ is
\begin{equation}
u^t = \frac{1}{\sqrt{y}},
\end{equation}
and therefore the only non-zero component of $a^\mu$ is
\begin{equation}
a^\lambda = u^tu^\lambda_{;t} = \Gamma^\lambda_{tt}(u^t)^2 = \frac{2}{y'}.
\end{equation}
As one may observe, $a^\mu$ is indeed perpendicular to our two-surface $S$. The norm of $a^\mu$ is:
\begin{equation}
a=\sqrt{a_\lambda a^\lambda} = \frac{1}{\sqrt{y}},
\end{equation}
which is exactly what one might have expected on grounds of Eqs. (5) and (8). The exterior curvature tensor on the two-surface $S$, in turn, is:
\begin{equation}
k_{jk} = -\frac{y'}{2\sqrt{y}}\Gamma^\lambda_{jk} = \frac{\sqrt{y}}{y'}{q'}_{jk},
\end{equation}
and its trace is:
\begin{equation}
k = q^{jk}k_{jk} = \frac{\sqrt{y}}{y'}q^{jk}{q'}_{jk}.
\end{equation}
In Eqs. (21) and (22) the prime means the derivative with respect to $\lambda$. Hence the Brown-York energy of Eq. (1) takes the form:
\begin{equation}
E_{BY} = -\frac{1}{8\pi}\oint_S\left(\frac{\sqrt{y}}{y'}q^{jk}{q'}_{jk} - k_0\right)\sqrt{q}\,d\chi^2\,d\chi^3,
\end{equation}
where we have integrated the coordinates $\chi^2$ and $\chi^3$ over the two-surface $S$. 

    Consider now what happens to the Brown-York energy, if we change the geometrical parameters $\alpha^s$ of spacetime and, at the same time, the location of the two-surface $S$ in such a way that the proper acceleration is kept as a constant. Since the proper acceleration $a$ depends on $y$ as in Eq. (20), $y$ must be kept as a constant. This means that if the parameters $\alpha^s$ take on small changes $d\alpha^s$, the coordinate $\lambda$ takes on the change $d\lambda$ such that:
\begin{equation}
dy = \frac{\partial y}{\partial\lambda}\,d\lambda + \frac{\partial y}{\partial\alpha^s}\,d\alpha^s = 0,
\end{equation}
which implies:
\begin{equation}
\frac{\partial y}{\partial\alpha^s}\,d\alpha^s = -y'\,d\lambda.
\end{equation}
Because $y$ depends on the parameters $\alpha^s$, but the metric tensor $q_{jk}$ does not, the change taken by the Brown-York energy is:
\begin{equation}
dE_{BY} = \frac{\partial E_{BY}}{\partial\alpha^s}\,d\alpha^s = -\frac{1}{8\pi}\left\lbrace\oint_S\left\lbrack\frac{1}{2\sqrt{y}y'}\frac{\partial y}{\partial\alpha^s} - \frac{\sqrt{y}}{(y')^2}\frac{\partial y'}{\partial\alpha^s}\right\rbrack q^{jk}{q'}_{jk}\sqrt{q}\,d\chi^2\,d\chi^3\right\rbrace\,d\alpha^s.
\end{equation}
At the horizon $y=0$, whereas $y'>0$. Hence we may ignore the second term inside of the brackets, when we are just outside of the horizon. Employing Eqs. (20) and (25) we may write $dE_{BY}$ in terms of the proper acceleration $a$ and the change $d\lambda$ in the coordinate $\lambda$ as:
\begin{equation}
dE_{BY} = \frac{a}{8\pi}\left(\oint_S\frac{1}{2}q^{jk}{q'}_{jk}\sqrt{q}\,d\chi^2\,d\chi^2\right)\,d\lambda,
\end{equation}
or:
\begin{equation}
dE_{BY} = \frac{a}{8\pi}\,dA,
\end{equation}
where
\begin{equation}
dA = \frac{dA}{d\lambda}\,d\lambda = \left\lbrack\oint_S\left(\frac{d}{d\lambda}\sqrt{q}\right)\,d\chi^2\,d\chi^3\right\rbrack\,d\lambda = \left(\oint_S\frac{1}{2}q^{jk}{q'}_{jk}\sqrt{q}\,d\chi^2\,d\chi^2\right)\,d\lambda,
\end{equation}
is the change in the area 
\begin{equation}
A = \oint_S\sqrt{q}\,d\chi^2\,d\chi^3
\end{equation}
of the two-surface $S$. Hence we have managed to derive Eq. (2) in a very general case.

As an example, consider the Schwarzschild spacetime equipped with the line element
\begin{equation}
ds^2 = -\left(1 - \frac{2M}{r}\right)\,dt^2 + \frac{dr^2}{1 - \frac{2M}{r}} + r^2\,d\theta^2 + r^2\sin^2(\theta)\,d\phi^2,
\end{equation}
where $M$ is the Schwarzshild mass. The only non-zero components of the exterior curvature tensor on the two-sphere, where $r=constant$ are:
\begin{subequations}
\begin{eqnarray}
k_{\theta\theta} &=& -\frac{1}{\sqrt{1 - \frac{2M}{r}}}\Gamma^r_{\theta\theta} = \sqrt{1 - \frac{2M}{r}}r,\\
k_{\phi\phi} &=& -\frac{1}{\sqrt{1 - \frac{2M}{r}}}\Gamma^r_{\phi\phi} = \sqrt{1 - \frac{2M}{r}}r\sin^2(\theta),
\end{eqnarray}
\end{subequations}
and therefore the Brown-York energy of Eq. (1) inside of the two-sphere takes the form:
\begin{equation}
E_{BY} = -\frac{1}{8\pi}\left(\sqrt{1 - \frac{2M}{r}}\frac{2}{r} - \frac{2}{r}\right)4\pi r^2 = -r\sqrt{1 - \frac{2M}{r}} + r.
\end{equation}
The only non-zero component of the proper acceleration vector field $a^\mu$ of an observer at rest on the two-sphere is:
\begin{equation}
a^r = u^tu^r_{;t} = \Gamma_{tt}^r(u^t)^2 = \frac{M}{r^2},
\end{equation}
and its norm is:
\begin{equation}
a = \sqrt{a_ra^r} = \left(1 - \frac{2M}{r}\right)^{-1/2}\frac{M}{r^2}.
\end{equation}
The geometrical parameter of spacetime is now the Schwarzschild mass $M$, whereas the coordinate $\lambda$ is represented by $r$. If $M$ and $r$ take on small changes $dM$ and $dr$ such that $a$ is kept as a constant, we must have:
\begin{equation}
da = \frac{\partial a}{\partial r}\,dr + \frac{\partial a}{\partial M}\,dM = 0,
\end{equation}
or:
\begin{equation}
dM = \frac{M^2 + \left(1 - \frac{2M}{r}\right)2Mr}{Mr + \left(1 - \frac{2M}{r}\right)r^2}\,dr.
\end{equation}
The resulting change in the Brown-York energy is:
\begin{equation}
dE_{BY} = \frac{\partial E_{BY}}{\partial M}\,dM = \left(1 - \frac{2M}{r}\right)^{-1/2} \frac{M^2 + \left(1 - \frac{2M}{r}\right)2Mr}{Mr + \left(1 - \frac{2M}{r}\right)r^2}\,dr,
\end{equation}
where we have used Eq. (37). Just outside of the horizon, where $r=2M$ we may write, in the leading approximation:
\begin{equation}
dE_{BY} = \left(1 - \frac{2M}{r}\right)^{-1/2}\frac{M}{r^2}r\,dr,
\end{equation}
and hence Eq. (35) implies:
\begin{equation}
dE_{BY} = \frac{a}{8\pi}\,dA,
\end{equation}
where $dA = 8\pi r\,dr$ is the change in the area of the two-sphere. The result is consistent with Eq. (2).

In Ref. \cite{seite1} it was shown that in the Reissner-Nordstr\"om spacetime the change $dE_{BY}$ in the Brown-York energy inside of a closed two-surface just outside of the event horizon of the Reissner-Nordstr\"om black hole equals with the amount of energy carried by the matter fields through that surface from the point of view of an observer at rest with respect to that surface. There are good grounds to believe the result to be of general validity. It would be interesting to study the possible relationship of this result with an interesting hypothesis published in Refs. \cite{ysi} and \cite{kymppi} that at the horizon of spacetime the sum of the gravitational and the non-gravitational energies equals zero. This topic, however, is very wide, and its proper investigation must therefore be left to another publication. 

In this paper we have managed to obtain  a very simple relationship between the change in the Brown-York energy inside of a closed two-surface just outside of a horizon of spacetime, and the change in the area of that two-surface. The essential point in our derivation was an assumption that the proper acceleration $a$ of an observer at rest with respect to the surface was kept as a constant during the process. If we keep the proper acceleration as a constant, the changes in the Brown-York energy reflect the changes in the true, physical properties of spacetime, rather than the changes in the state of motion of the observer. Our result suggests that from the point of view of our observer the energy operator in curved  spacetime takes an extremely simple form:
\begin{equation}
\hat{H} = \frac{a}{8\pi}\hat{A}.
\end{equation}
Since our two-surface is just outside of the horizon, we may identify the operator $\hat{A}$ as the area operator of the horizon, and  it depends on the model of quantum gravity. Loop quantum gravity-related consequences of Eq. (41) in black hole thermodynamics and cosmology have been considered, for instance, in Ref. \cite{nee}.

\end{document}